\documentclass[twocolumn,aps,prl,floats,showpacs]{revtex4}

\renewcommand{\vr}{\mathbf{r}}

\usepackage{bm}
\usepackage{dcolumn}
\usepackage{epsfig}

\begin{document}

\title{Interplay of Ehrenfest time and dephasing time in ballistic conductors}

\author{Alexander Altland$^1$, Piet W.\ Brouwer$^2$, and Chushun Tian$^1$,}

\affiliation{$^1$Institut f{\"u}r Theoretische Physik,
Universit{\"a}t zu K{\"o}ln, K{\"o}ln, 50937, Germany \\
$^2$ Laboratory of Atomic and Solid State Physics, Cornell
  University, Ithaca 14853, USA}

\date{\today}

\pacs{73.23.-b, 05.45.Mt, 73.20.Fz}

\begin{abstract}
Quantum interference corrections in ballistic conductors require a minimal time: the Ehrenfest time. In this letter, we investigate the fate of the interference corrections to quantum transport in bulk ballistic conductors if the Ehrenfest time and the dephasing time are comparable.
\end{abstract}

\maketitle

{\em Introduction.}
In recent years, the Ehrenfest time $\tau_{\rm E}$ has been recognized as a time scale of profound relevance to the physics of systems interfacial between the mesoscopic and the nanoscopic regime \cite{kn:aleiner1996}.
Loosely speaking, $\tau_{\rm E}$ is the time it takes before a minimal wave packet propagating in a chaotic background looses its integrity and spreads over scales of classical proportions \cite{kn:larkin1968,kn:zaslavsky1981}. Therefore 
(i) the Ehrenfest time defines a time threshold before the wave nature of electrons begins to modify the classical behavior of observable system properties. Accordingly, 
(ii) there is a general expectation that quantum effects are multiplied by exponential weighting factors $\exp(- \alpha \tau_{\rm E}/t_0)$, where $t_0$ is the (smallest) characteristic time scale of the quantum effect. 
This expectation has been confirmed for the Ehrenfest-time related suppression of weak localization \cite{kn:aleiner1996,kn:adagideli2003,kn:rahav2005,kn:jacquod2006} and shot noise \cite{kn:agam2000,kn:whitney2006} in chaotic quantum dots, with $t_0$ taken to be the dot's mean dwell time $\tau_{\rm D}$, or Ehrenfest-oscillations of the weak localization corrections to the ac conductivity of a random collection of antidots \cite{kn:aleiner1996} and time-dependent diffusion in periodically kicked atomic gases \cite{kn:tian2004}, with $t_0 = i \omega^{-1}$ taken to be the inverse angular frequency .

In this letter, we consider the competition between the Ehrenfest time and the dephasing time $\tau_{\phi}$. Whereas $\tau_{\rm E}$ is the minimal time needed for quantum interference, $\tau_{\phi}$ sets the long-time cut-off for interference processes. The competition between $\tau_{\rm E}$ and $\tau_{\phi}$ is particularly relevant for quantum corrections in bulk conductors, for which the dwell time $\tau_{\rm D}$ has no significance. In particular, we'll address the question whether one may expect a suppression of quantum corrections proportional to $\exp(- \alpha \tau_{\rm E}/\tau_{\phi})$, according to the general expectation (ii) mentioned above. In a subtle manner, the answer depends on whether the dephasing originates from electron-electron interactions or from an external source (such as applied microwave radiation). Conceptually, the observation of an Ehrenfest-time dependence of quantum interference corrections to the conductance has exponential sensitivity to the microscopic mechanism of dephasing.

To date, there are only a few experimental signatures of the Ehrenfest time. Oberholzer {\em et al.} found a $\tau_{\rm E}$-related suppression of the shot noise of a chaotic cavity upon decreasing $\tau_{\rm D}$ \cite{kn:oberholzer2002}. Shot noise, however, is insensitive to the presence of dephasing. Yevtushenko {\em et al.}\ observed an exponential suppression of weak localization in an antidot lattice with increasing temperature $T$ and attributed this observation to the competition of $\tau_{\rm E}$ and $\tau_{\phi}$ \cite{kn:yevtushenko2000}. The theoretical insights reported here should be relevant for the interpretation of the latter experiment.

\begin{figure}
\begin{center}
\leavevmode \epsfxsize=8cm \epsfbox{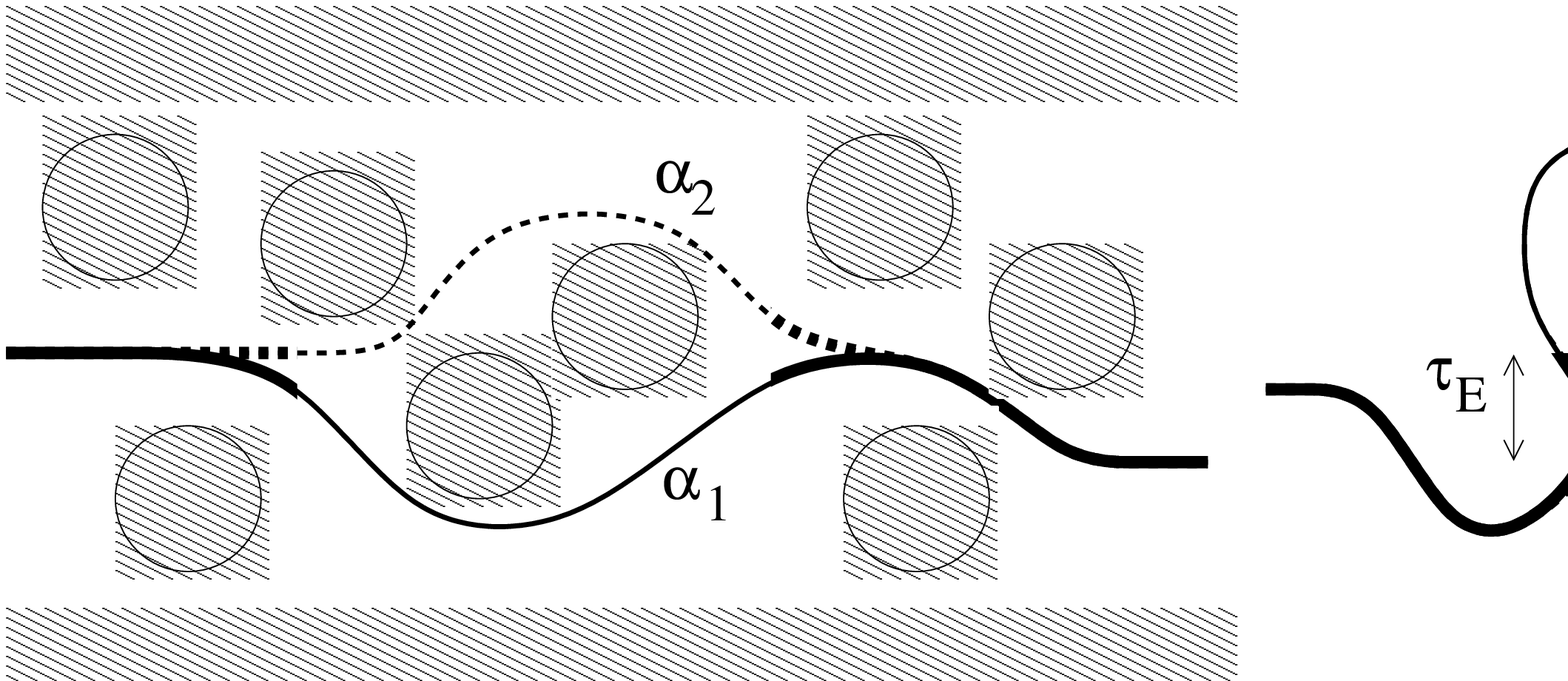}
\end{center}
\caption{Schematic drawing of a Lorentz gas and a generic pair of trajectories $\alpha_1$ (solid) and $\alpha_2$ (dotted) that contributes to the quantum interference correction $\delta G$ to the conductance (left) and of a trajectory pair that contributes to the ensemble average $\langle \delta G \rangle$ (right). The `entrance' and `exit segments' are shown thick; the central segment is shown thin. The Lyapunov region is indicated by the double arrow.}
  \label{fig:1}
\end{figure}

{\em Semiclassical picture.}
We focus our discussion on a ballistic conductor in which the large-scale electron dynamics is diffusive, such as the Lorentz gas, a random collection of disc-like scatterers, see Fig.\ \ref{fig:1}. We employ semiclassical language wherein the conductance $G$ in the absence of electron-electron interactions is expressed as a sum over pairs of classical trajectories $\alpha_1$ and $\alpha_2$ \cite{kn:jalabert1990},
\begin{equation}
  G = \frac{2 e^2}{h}
  \int dp_{\perp} dp_{\perp}' \sum_{\alpha_1,\alpha_2} 
  A_{\alpha_1}^{\vphantom{*}} A_{\alpha_2}^* .
  \label{eq:Gsemi}
\end{equation}
These originate from the source lead with momentum component $p_{\perp}$ perpendicular to the lead axis and end at the drain lead with perpendicular momentum component $p_{\perp}'$. In Eq.\ (\ref{eq:Gsemi}), $A_{\alpha}$ is the quantum mechanical transition amplitude of trajectory $\alpha$. Both $\alpha_1$ and $\alpha_2$ {\em exit} the system at the same time $t$, whereas the entrance times are different if the durations $t_{\alpha_1}$ and $t_{\alpha_2}$ of the two trajectories differ. The diagonal part of the sum corresponds to the classical (Drude) conductance; the remaining part of the summation, which is over pairs of different trajectories $\alpha_1 \neq \alpha_2$, is the quantum correction $\delta G$ to the conductance. The ensemble average $\langle \delta G \rangle$ is the weak localization correction to the conductance.

The large parameter justifying the semiclassical formulation is the ratio of classical, macroscopic length scales, such as radius $R$ of the scattering discs or their mean distance, and the Fermi wavelength $\lambda_F$. In our discussion of dephasing we assume that the dephasing length $l_{\phi}$ is also macroscopic: If not, the electronic phase is destroyed before even the smallest interference loop can be formed, and no quantum interference corrections can exist.

A nonzero contribution to $\delta G$ --- even before ensemble averaging --- exists only for those pairs of trajectories for which not only the entrance and exit momenta,
but also the entrance and exit positions 
are equal, up to a quantum uncertainty \cite{kn:aleiner1996}. Hence a typical pair of classical trajectories contributing to $\delta G$ is as shown in Fig.\ \ref{fig:1}a: The trajectories originate from positions a quantum uncertainty apart and diverge exponentially by virtue of the chaotic classical dynamics. When the distance between the two trajectories exceeds a distance $L_{\rm c} \sim R$, their classical motion has become uncorrelated. Finally, the trajectories join again and reach the exit contact at positions only a quantum distance apart. 


The generic situation shown in Fig.\ \ref{fig:1}a gives a contribution to the sample-specific quantum correction $\delta G$ but not to its ensemble average. In order to have a contribution to $\langle \delta G \rangle$, the two trajectories $\alpha_1$ and $\alpha_2$ should be piecewise equal. This is achieved if the trajectories have a small-angle self encounter, as shown schematically in the right panel of Fig.\ \ref{fig:1} \cite{kn:aleiner1996,kn:richter2002}. The duration of the encounter or `Lyapunov region', measured as the time during which the separation between the two trajectories is below the classical cut-off $L_{\rm c}$ is the Ehrenfest time $\tau_{\rm E} = \lambda^{-1} \ln(L_{\rm c}/\lambda_F)$, where $\lambda$ is the Lyapunov exponent driving the separation of initially close trajectories up to classical separations of order $L_{\rm c}$. 

\begin{figure}
\begin{center}
\leavevmode \epsfxsize=7cm \epsfbox{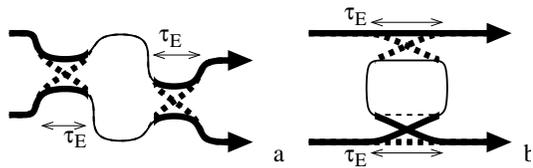}
\end{center}
\caption{Semiclassical representation of the two distinct
    contributions to the conductance fluctuations. The entrance and exit segments are shown thick, the central segment is shown thin.}
  \label{fig:2}
\end{figure}

The magnitude of the sample-specific quantum correction $\delta G$ is measured through the conductance variance, $\mbox{var}\, G = \langle \delta G^2 \rangle - \langle \delta G \rangle^2$. Since the square of the conductance is expressed as a quadruple sum over classical trajectories, one needs to identify two pairs of trajectories of the type shown in Fig.\ \ref{fig:1}a such that the product of all four transition amplitudes is a weakly fluctuating quantity. Two topologically distinct contributions of this type exist \cite{kn:altshuler1986}, see Fig.\ \ref{fig:2} \cite{foot}. In the first of these, two trajectories initially a quantum distance apart and entering the first factor $\delta G$ split to join with the quantum amplitude of two trajectories of the second factor $\delta G$. In the second contribution, the two trajectories in a pair entering into the same factor $\delta G$ differ by a `loop', which one trajectory travels through and the other does not. The two pairs are arranged such that the same loop is traversed in both cases. We note that these two contributions can be linked to the two primary contributions to the universal conductance fluctuations in standard disordered conductors \cite{kn:altshuler1986}. The first contribution corresponds to the contribution of fluctuations of the diffusion constant, whereas the second contribution represents the density of states contribution to the conductance fluctuations.

The presence of a time-dependent potential, either from an intrinsic source, such as electron-electron interactions, or from an external source, may change the phases of the amplitudes $A_{\alpha_1}$ and $A_{\alpha_2}$ in different ways. Such dephasing causes a suppression of the quantum interference correction. The central question of this letter is whether dephasing can occur during the Lyapunov regions. Only dephasing in the Lyapunov regions can give rise to an exponential dependence $\propto \exp(-\alpha t_{\rm E}/\tau_{\phi})$ of the quantum corrections. Dephasing outside the Lyapunov regions is described by the standard theory of disordered conductors \cite{kn:altshuler1985a,kn:aleiner2002b,kn:narozhny2002}.

For the purpose of addressing the role of dephasing, we separate a generic trajectory pair of Fig.\ \ref{fig:1}a into three parts: an `exit segment', consisting of the stretch of correlated propagation of both trajectories near the exit contact, an `entrance segment', consisting of the stretch of correlated propagation of both trajectories near the entrance contact (without parts of the trajectories that are already included in the exit segment), and the remaining `central segment'. With this definition, the Lyapunov regions are part of the exit and entrance segments. We note that the central segment may be empty for one of the trajectories in a pair. This is the case for the trajectories shown in Fig.\ \ref{fig:2}b. 

{\em External source of dephasing.} 
Although the frequency $\omega$ and wavenumber $q$ of the dephasing potential $V(\vr,t)$ can be controlled arbitrarily, in most cases of practical interests --- a fluctuating gate potential or an applied microwave field \cite{kn:vavilov1999} --- the relevant wavenumbers are macroscopic, not microscopic. Macroscopic fluctuating potentials cause forward scattering only: The phase accumulated by the electrons is changed, but the electrons continue to propagate along classical trajectories. This implies that the semiclassical picture underlying Eq.\ (\ref{eq:Gsemi}) can still be used; The trajectories $\alpha_1$ and $\alpha_2$ acquire a phase shift $\phi(\alpha_{1,2},t)$ which depends on the time $t$ the trajectories reach the exit lead,
\begin{equation}
  \phi(\alpha,t) = \hbar^{-1} \int_{t-t_{\alpha}}^{t}
  dt' V(\vr_{\alpha}(t'),t').
\end{equation}
The accumulated phase $\phi = \phi_{\rm entr} + \phi_{\rm centr} + \phi_{\rm exit}$ can be separated into contributions from the entrance, central, and exit segments. In the entrance and exit segments, $\alpha_1$ and $\alpha_2$ see the same fluctuating potential. This implies $\phi_{\rm exit}(\alpha_1,t) = \phi_{\rm exit}(\alpha_2,t)$. The phase shifts $\phi_{\rm entr}(\alpha_1,t)$ and $\phi_{\rm entr}(\alpha_2,t)$ may be different, however, if $\alpha_1$ and $\alpha_2$ have different durations $t_{\alpha_1}$ and $t_{\alpha_2}$.

The trajectory pairs contributing to the ensemble average $\langle \delta G \rangle$ have $t_{\alpha_1} = t_{\alpha_2}$. Hence, there is no dephasing in either exit or entrance regions. Since these contain the Lyapunov region, we conclude that for weak localization there is no exponential suppression $\propto \exp(\alpha \tau_{\rm E}/\tau_{\phi})$. Alternatively, one notes that the situation in the entrance and exit segments is effectively one dimensional, and that the only effect of a forward scattering potential in one dimension is to change the distribution function. Such a change is inconsequential for weak localization, which is independent of the distribution function.

The situation is more complicated for the conductance fluctuations. Here the two trajectories $\alpha_1$ and $\alpha_2$ can have different durations. As a result, one finds a suppression of the conductance fluctuations that is proportional to $\exp(-t_{\rm entr}/t_0)$, where $t_{\rm entr}$ is the duration of the `entrance segment' and 
\begin{eqnarray}
  t_0^{-1} &=& \frac{1}{2 t_{\rm entr}}
  \overline{[\phi(\alpha_1) - \phi(\alpha_2)]^2}.
  \nonumber \\ & \sim &
  \frac{1}{\tau_{\phi}} 
  \left[ 1 - \cos(\omega (t_{\alpha_1} - t_{\alpha_2})\right],
\end{eqnarray}
where $\tau_{\phi}$ is the dephasing time in the central segment, in which the trajectories $\alpha_1$ and $\alpha_2$ are uncorrelated. The bar denotes an average along the entrance segment of the classical trajectory. The difference $t_{\alpha_1} - t_{\alpha_2}$ is typically of order $\min(T^{-1},\tau_{\rm D})$. Whereas an exponential suppression proportional to $\exp(-t_{\rm entr}/t_0)$ implies an exponential suppression $\propto \exp(-\tau_{\rm E}/t_0)$ for the contribution to $\mbox{var}\, G$ of Fig.\ \ref{fig:2}a, it does not imply a similar exponential suppression for the contribution of Fig.\ \ref{fig:2}b: In  Fig.\ \ref{fig:2}b the entrance segment contains no Lyapunov region. Hence, we conclude that $\mbox{var}\, G$ is not exponentially suppressed at large $\tau_{\rm E}/\tau_{\phi}$ for an external source of dephasing.


{\em Dephasing from electron-electron interactions.} Even without external sources, the electrons will be subject to a time-dependent fluctuating potential: the potential generated by the other electrons through electron-electron interactions. Now the characteristic wavenumber and frequency are not controlled externally, but they are set by the temperature and the electron dynamics. In a disordered conductor in one or two dimensions, the dephasing rate $\tau_{\phi}^{-1}$ can be written as a sum of two contributions \cite{kn:narozhny2002},
\begin{equation}
  \tau_{\phi}^{-1} = \tau_{\phi,{\rm diff}}^{-1} + \tau_{\phi,{\rm ball}}^{-1},
  \label{eq:diffball}
\end{equation}
where $\tau_{\phi,{\rm diff}}$ and $\tau_{\phi,{\rm ball}}$ represent the dephasing times from time-dependent fluctuations of the interaction potential $V(\vr,t)$ on length scales above and below the mean free path $l = v_F \tau$, respectively. At low temperatures the first term in Eq.\ (\ref{eq:diffball}) dominates the dephasing rate \cite{kn:altshuler1982b}, whereas the second term dominates at high temperatures. The two contributions are comparable for $T \sim \hbar/\tau$ \cite{kn:narozhny2002}.

As in the case of dephasing from an external source, a time-dependent potential with a macroscopic spatial dependence changes the electronic phase, but not the classical trajectories. Repeating the arguments above, we thus conclude that there is no additional contribution to dephasing from the exit segment. However, there is no contribution to dephasing from the entrance segment either. This is different from the case of external dephasing. The reason is that the forward scattering potential is generated by electrons in equilibrium, so that the fluctuating interaction potential does not broaden the distribution function in the entrance segment. Therefore, the dephasing processes that enter into $\tau_{\phi,{\rm diff}}$ do not lead to any $\tau_{\rm E}$ dependence of weak localization or the conductance fluctuations. (The absence of a suppression $\propto e^{-\alpha \tau_{\rm E}/\tau_{\phi,{\rm diff}}}$ for weak localization was already noted in Ref.\ \onlinecite{kn:aleiner1996}.)

The situation is different for the ballistic contribution to dephasing. Although the effect of a time-dependent potential that varies on sub-macroscopic length scales can not be described quantitatively using the semiclassical picture of Eq.\ (\ref{eq:Gsemi}) because such a potential may change both trajectories and phases, the impact of the fluctuating potential during the Lyapunov regions can be argued using qualitative arguments. The distance $d$ between the two trajectories in the first Lyapunov regions of Fig.\ \ref{fig:1}b or Fig.\ \ref{fig:2} is estimated as
\begin{equation}
  d \sim \lambda_F e^{\lambda t},
\end{equation}
where $t$ is the time measured since entry of the Lyapunov region. Upon exit of the Lyapunov region, after a time $\tau_{\rm E}$, one has $d \sim R$. The same estimate holds for the second Lyapunov region provided $t$ is interpreted as the time before exit. In a Lyapunov region, phase breaking from potential fluctuations with wavenumber $q$ can occur only if $d q \gtrsim 1$ (see also Ref.\ \onlinecite{kn:petitjean2007}). Repeating the calculation of the ballistic dephasing rate $\gamma_{\phi,{\rm ball}}$ \cite{kn:narozhny2002} with the condition $q \gtrsim 1/d$, one finds that the dephasing rate is modified as
\begin{equation}
  \tau_{\phi,{\rm ball}}(d) = 
  \tau_{\phi,{\rm ball}} \frac{\ln (R/\lambda_F)}{\ln (d/\lambda_F)}, 
  \label{eq:tauphid}
\end{equation}
where $\tau_{\phi,{\rm ball}}$ is the dephasing time outside the Lyapunov region. Averaging over the Lyapunov region, one arrives at an effective dephasing rate for weak localization that is only half of the ballistic dephasing rate outside the Lyapunov region. Since electrons contributing to weak localization pass through the same Lyapunov region twice, we conclude that 
\begin{equation}
  \langle \delta G \rangle \propto \exp(-\tau_{\rm E}/\tau_{\phi,{\rm ball}}).
  \label{eq:dGtauphi}
\end{equation}
For universal conductance fluctuations the contribution of Fig.\ \ref{fig:2}b decays slower with increasing $\tau_{\rm E}$ than the contribution of Fig.\ \ref{fig:2}a: For Fig.\ \ref{fig:2}b one can judiciously pair the two interfering trajectories $\alpha_1$ and $\alpha_2$ such that their distance is never larger than $(R \lambda_F)^{1/2}$, while in Fig.\ \ref{fig:2}a the distance between trajectories in the Lyapunov regions can be as large as $R$. The effective dephasing rate, averaged over the duration of the Lyapunov region, then becomes $1/4  \times \tau_{\phi,{\rm ball}}^{-1}$, so that
\begin{equation}
  \mbox{var}\, G \propto \exp(-\tau_{\rm E}/2 \tau_{\phi,{\rm ball}})
  \label{eq:varGtauphi}
\end{equation}
if $\tau_{\rm E} \gg \tau_{\phi,{\rm ball}}$. 

{\em Ballistic quantum dots.}
These ideas also apply to the Ehrenfest-time dependence of weak localization and conductance fluctuations in a ballistic quantum dot. For external dephasing with a macroscopic spatial dependence of the dephasing potential, an exact calculation along the lines of Ref.\ \onlinecite{kn:brouwer2006} shows that both $\langle \delta G \rangle$ and $\langle \mbox{var}\, G \rangle$ are independent of $\tau_{\rm E}/\tau_{\phi}$. For intrinsic dephasing one expects that Eqs.\ (\ref{eq:dGtauphi}) and (\ref{eq:varGtauphi}) remain valid, with $\tau_{\phi,{\rm ball}}$ replaced by the total dephasing time $\tau_{\phi}$, because $\tau_{\phi,{\rm diff}}^{-1} = 0$ in a ballistic quantum dot.

{\em Discussion.} 
We note that Eq.\ (\ref{eq:dGtauphi}) appeared previously in the literature, but for rather different reasons. Aleiner and Larkin \cite{kn:aleiner1996} arrive at an effective ballistic dephasing rate that is half the dephasing rate $\tau_{\phi,{\rm ball}}^{-1}$ outside the Lyapunov regions by artificially setting the dephasing rate to zero in the first half of each Lyapunov region. Petitjean {\em et al.}\ find Eq.\ (\ref{eq:dGtauphi}), with $\tau_{\phi,{\rm ball}}$ replaced by $\tau_{\phi}$, for a quantum dot in which dephasing arises from a voltage probe ballistically coupled to the dot \cite{kn:petitjean2007}.
Tworzydlo {\em et al.}, who considered a tunnel-coupled voltage probe,
reported $\langle \delta G \rangle \propto \exp(-\tau_{\rm E}/\tau_{\phi})$ based on an `effective random matrix theory' which neglects the second passage through the Lyapunov region \cite{kn:tworzydlo2004c}. The correct result for dephasing from a tunnel-coupled voltage probe is $\langle \delta G  \rangle \propto \exp(-2 \tau_{\rm E}/\tau_{\phi})$ \cite{kn:brouwer2006,kn:petitjean2007}. For the variance of the conductance, Ref.\ \onlinecite{kn:tworzydlo2004c} finds $\mbox{var}\, G \propto \exp(-2 \tau_{\rm E}/\tau_{\phi})$, which is the correct result for the model employed there.

The only experiment to date that claims to have observed the $\tau_{\rm E}$ dependence of weak localization, Ref.\ \cite{kn:yevtushenko2000}, derives this claim from the observed exponential temperature dependence of the weak localization correction for a two-dimensional collection of randomly placed antidots. Reference \cite{kn:yevtushenko2000} used Eq.\ (\ref{eq:dGtauphi}), but with $\tau_{\phi,{\rm ball}}$ replaced by $\tau_{\phi,{\rm diff}}$, to analyze their data. Since $\tau_{\phi,{\rm diff}} \propto T^{-1}$ in two dimensions, this would indeed explain the observed temperature dependence of $\langle \delta G \rangle$. The correct $\tau_{\rm E}$-dependence of the weak localization correction involves the ballistic dephasing time $\tau_{\phi,{\rm ball}}$, however, which is proportional to $T^{-2}$, not $T^{-1}$. Further complications arise because a large part of the experimental data is for $T \sim \hbar/\tau$, for which neither $\tau_{\phi,{\rm ball}}$ nor $\tau_{\phi,{\rm diff}}$ dominates the dephasing rate. This rules out an unambiguous identification of the role of the Ehrenfest time from the observed temperature dependence of weak localization alone.

We are grateful to I.~Aleiner, A.~Andreev, C.~Beenakker,
A.~Kamenev, R.~Whitney, S.~Rahav, and M.~Vavilov for useful discussions. 
C.~T. acknowledges the hospitality of CASTU at 
Beijing, where this work was initiated. This work was supported by
Transregio SFB 12 of the Deutsche Forschungsgemeinschaft, by the
Packard Foundation, and by the NSF under grant no.\ 0334499.
After completion of this manuscript, Ref.\ \onlinecite{kn:petitjean2007}
appeared on the cond-mat archive, in which similar results were obtained
with regard to dephasing from an external source.


\end{document}